\documentclass[10pt]{article}

\usepackage[english]{babel}

\usepackage{epsfig}
\usepackage{dcolumn}
\usepackage{amsmath}
\usepackage{changebar}
\usepackage{graphicx}

\newcommand{\ba} {\begin{eqnarray}}
\newcommand{\ea} {\end{eqnarray}}
\usepackage{graphicx}
\usepackage{dcolumn} 
\usepackage{bm}      
\usepackage{umlaut}
\begin{document}
\renewcommand{\figurename}{Fig.}
\renewcommand{\tablename}{Tab.}
\title{Event-by-Event Fluctuations of Particle Ratios in Heavy-Ion Collisions}

\author{ A.~Tawfik\thanks{tawfik@physik.uni-bielefeld.de} \\
 {\small Hiroshima University, 1-7-1 Kagami-yama, Higashi-Hiroshima 
 Japan }  
}

\date{}
\maketitle

\begin{abstract}
We study event-by-event dynamical fluctuations of various particle
 ratios at different energies. We assume that particle production in
 final state is due to chemical equilibrium processes. We compare results from
 resonance  gas model with available experimental data. At SPS energies,
 the model can very well reproduce the experimentally measured
 fluctuations. We make predictions for dynamical fluctuations of
 strangeness and  non-strangeness particle ratios. We found that the
 energy-dependence is 
 non-monotonic. Furthermore, we found that fluctuations strongly depend
 on  particle ratios.    
\end{abstract}

\section{\label{sec:A}Introduction}
Understanding dynamical properties of hot and dense matter is a key question in
heavy-ion collisions experiments. Event-by-event fluctuations have been
suggested~\cite{shury,rajag,koch} to provide comprehensive
characteristics of this matter. They are crucial observations to check
hypothesis of chemical equilibrium in heavy-ion
collisions~\cite{giorg}. Event-by-event fluctuations of very few
particle ratios have been studied in several experiments at SPS and
RHIC energies~\cite{NA491,NA492,STAR}. In this letter, we purpose to study 
energy-dependence of event-by-event fluctuations in hadron resonance gas
model. Hypothetical phase transition to and from quark-gluon plasma can
be characterized by large fluctuations in particle 
yields~\cite{shury,rajag,koch}, which are accompanied by volume
fluctuations. The latter can be eliminated, when particle ratios are
considered~\cite{koch}. In this letter, we try to give answers to the
questions, whether strangeness quarks enhance dynamical fluctuations and
whether critical endpoint can be localized by means of dynamical
fluctuations.  Since experimental measurements of event-by-event
fluctuations are very much limited, we hope that our predictions
encourage specifying certain particle ratios and measure their
dynamical fluctuations. In this letter, we make predictions for dynamical
and statistical fluctuations of different particle ratios in dependence
on energy. The values of fluctuations strongly depend on particle
ratios. In some particle ratios, dynamical fluctuations are smaller
than statistical ones. In other particle ratios, dynamical fluctuations
are slightly greater than statistical ones. We also found that
fluctuations of hybrid and cascade baryons are dominant. All these
predictions are phenomenologically of great interest.

\section{\label{sec:2}Model}

Pressure in hadronic phase is given by contributions from all hadron
resonances treated as a free
gas~\cite{Karsch:2003vd,Karsch:2003zq,Redlich:2004gp,Tawfik:2004sw}. The
resulting pressure accounts for free as well as for strong 
interactions between resonances. In previous work~\cite{Tawfik:2004sw},
we have proved that the thermodynamics of strongly interacting system of
hadron resonances can be approximated by an ideal gas of stable and
resonance hadrons.

At
finite temperature $T$, strangeness $\mu_S$ and iso-spin  $\mu_{I_3}$
and baryo-chemical potential $\mu_B$, pressure of one 
hadron  reads
\begin{eqnarray}
\label{eq:lnz1} 
p(T,\mu_B,\mu_S,\mu_{I_3}) &=& \frac{g}{2\pi^2}T \int_{0}^{\infty}
           k^2 dk  \ln\left[1 \pm\,
           \gamma\,
           \lambda_B \lambda_S \lambda_{I_3}
	   e^{\frac{-\varepsilon(k)}{T}}\right], 
\end{eqnarray}
where $\varepsilon(k)=(k^2+m^2)^{1/2}$ is single-particle energy and $\pm$
stands for bosons and fermions, respectively. $g$ is spin-isospin
degeneracy factor. $\gamma\equiv\gamma_q^n\gamma_s^m$  are quark phase
space occupancy 
parameters, where $n$ and $m$ being number of light and strange quarks,
respectively. In this letter, we explicitly use the equilibrium value
for $\gamma$, the
unity. $\lambda=\exp(\mu/T)$ is fugacity, where $\mu$ is chemical
potential multiplied by corresponding charge. 

Quark chemistry is given by relating {\it hadronic} chemical potentials
to quark constituents. $\mu_B=3\mu_q$ and $\mu_S=\mu_q-\mu_s$, where $q$
and $s$ being light and strange quark quantum number, respectively. The
baryo-chemical potential for light quarks is
$\mu_q=(\mu_u+\mu_d)/2$. $\mu_S$ is calculated as a function of $T$ and
$\mu_B$ under the condition of strangeness conservation. Iso-spin
chemical potential $\mu_{I_3}=(\mu_u-\mu_d)/2$.

Particle number density is given by derivative of partition function
in Eq.~\ref{eq:lnz1} with 
respect to the chemical potential of interest. Fluctuations in
particle number are given by susceptibility density, which is second
derivative with respect to chemical potential. 
\ba \label{eq:n1} 
\langle n\rangle &=& \sum_i \frac{g_i}{2\pi^2} \int dk k^2
\frac{e^{(\mu_i-\varepsilon_i)/T}}{1\pm e^{(\mu_i-\varepsilon_i)/T}}, \\
\label{eq:dn1} 
\langle (\Delta n)^2\rangle &=& \sum_i \frac{g_i}{2\pi^2} \int dk k^2
           \frac{e^{(\varepsilon_i-\mu_i)/T}}
	   {\left(e^{(\varepsilon_i-\mu_i)/T}\pm1\right)^2} 
       =  \sum_i \frac{g_i}{2\pi^2} \int dk k^2  
           \frac{\langle n_i\rangle}{1\pm e^{(\mu_i-\varepsilon_i)/T}}
\ea

After freeze out, hadron resonances decay either to stable particles or
to other resonances. Particle number and fluctuation density in final state
have to take into account this chemical process.
\ba \label{eq:n2}
\langle n_i^{final}\rangle &=& \langle n_i^{direct}\rangle + \sum_{j\neq
i} b_{j\rightarrow i} \langle n_j\rangle,\\ \label{eq:dn2} 
\langle (\Delta n_{j\rightarrow i})^2\rangle &=& b_{j\rightarrow i} (1-b_{j\rightarrow
i}) \langle n_j\rangle + b_{j\rightarrow i}^2 \langle (\Delta
n_{j})^2\rangle 
\ea
where $b_{j\rightarrow i}$ being branching ratio for decay of $j$-th resonance
to $i$-th particle. Chemical freeze out is characterized by
$s/T^3$~\cite{Taw3}, where $s$ is the entropy density.

Fluctuations of particle ratio $n_1/n_2$ are~\cite{koch}
\ba  \label{eq:sigma}
\sigma^2_{n_1/n_2} &=& \frac{\langle (\Delta n_1)^2\rangle}{\langle n_1\rangle^2} + 
                       \frac{\langle (\Delta n_2)^2\rangle}{\langle n_2\rangle^2} - 
                     2 \frac{\langle \Delta n_1 \; \Delta
		     n_2\rangle}{\langle n_1\rangle \; \langle
		     n_2\rangle} 
\ea
which include dynamical as well as statistical fluctuations. Third term of
Eq.~\ref{eq:sigma} counts for fluctuations from hadron resonances that
decay into particle $1$ and particle $2$, simultaneously. In such a
mixing channel, all correlations including quantum statistics ones are
taken into account.  Obviously, 
this decay channel results in strong correlated particles. To extract
statistical fluctuation, we apply Poisson scaling in mixed decay
channels~\footnote[1]{Experimentally, there are various methods to
construct statistical fluctuations~\cite{STAR}. Frequently used
method is the one that measures particle ratios from mixing events.},
\ba \label{eq:sigmaStat}
(\sigma^2_{n_1/n_2})_{stat} &=& \frac{1}{\langle n_1\rangle} +
\frac{1}{\langle n_2\rangle} 
\ea
Subtracting Eq.~\ref{eq:sigmaStat} from Eq.~\ref{eq:sigma}, we get
dynamical fluctuations of particle ratio $n_1/n_2$. 
\ba
 \label{eq:sigma2}
(\sigma^2_{n_1/n_2})_{dyn} &=& 
          \frac{\langle n_1^2\rangle}{\langle n_1\rangle^2} +
          \frac{\langle n_2^2\rangle}{\langle n_2\rangle^2} -
         \frac{\langle n_1\rangle+\langle n_2\rangle +
	 2\langle n_1n_2\rangle}{\langle n_1\rangle\langle n_2\rangle}
\ea

\section{\label{sec:3}Results}

\begin{figure}[thb]
\centerline{
\includegraphics[width=7.5cm]{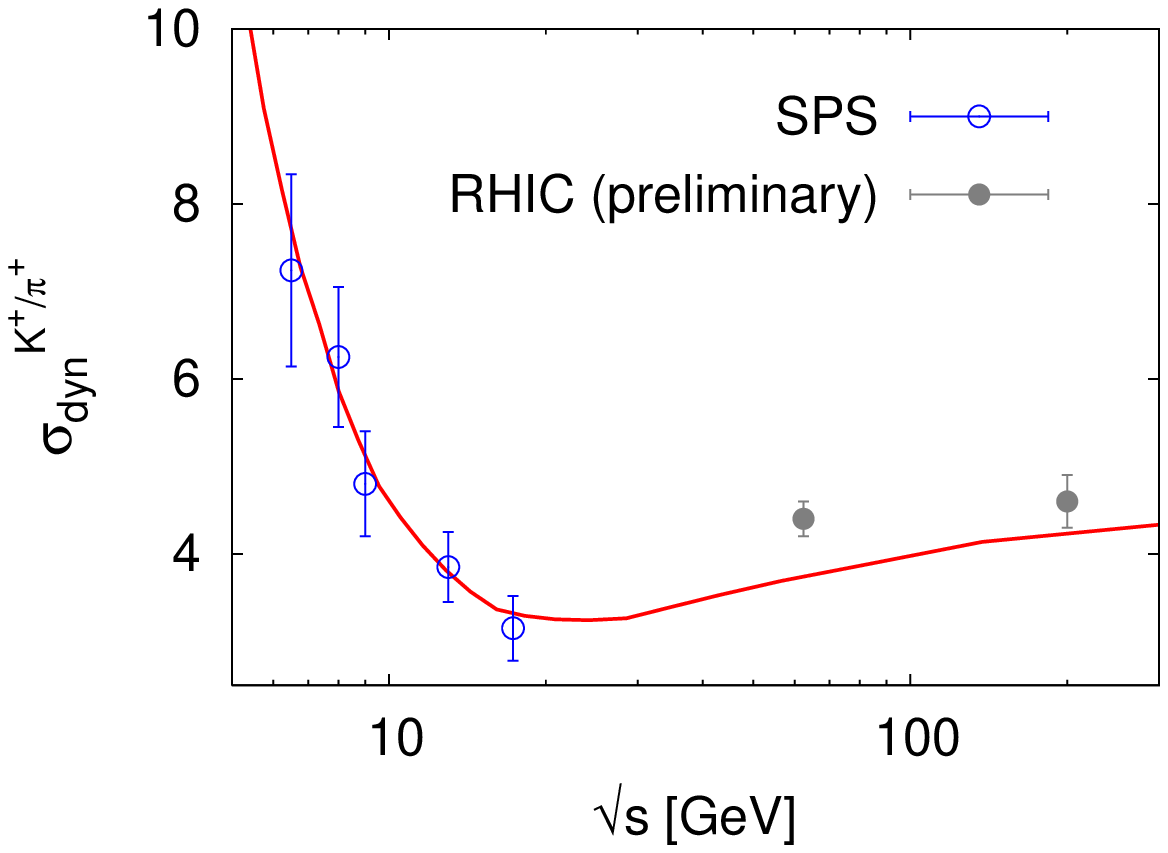}
\includegraphics[width=7.5cm]{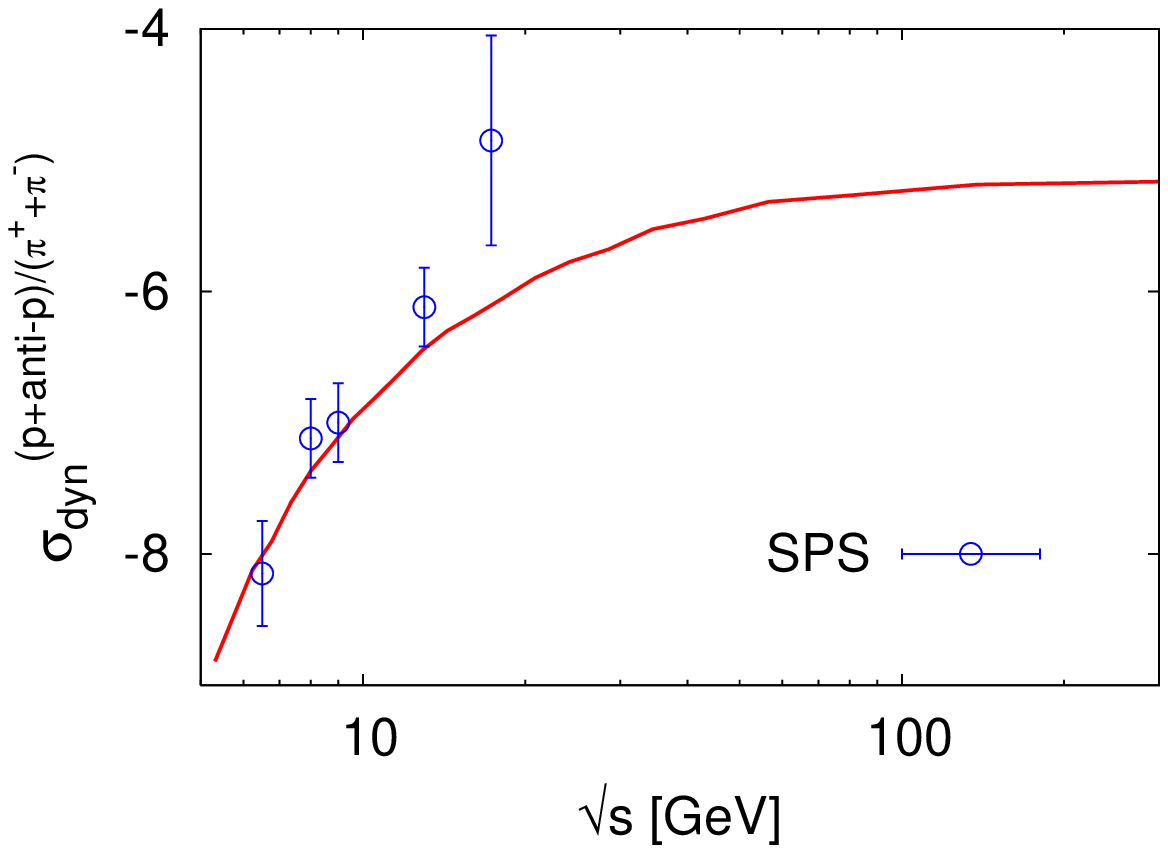}
}
\caption{Left panel: dynamical fluctuations of $K^+/\pi^+$ ratio as
 a function of square root of center of mass energy (curve) compared with
 experimental 
 results (circles). Results from SPS experiments~\cite{NA492} are drawn in open
 symbols. RHIC results are still preliminary (solid circles). Right
 panel shows dynamical fluctuations of  
 $(p+\bar{p})/(\pi^++\pi^-)$ ratios. Negative values are an indication to 
 dynamical fluctuations that are smaller than 
 statistical ones.} \label{Fig:kpi}  
\end{figure}

For the first time, experimental measurements of dynamical
fluctuations of particle ratios are systematically confronted
with theoretical predictions in Fig.~\ref{Fig:kpi}. There was an
earlier attempt to compare with preliminary results for
$K/\pi$ reported in~\cite{koch}. Is has been found that theoretical and
experimental ratios of dynamical to statistical fluctuations are
compatible with each other. Individual fluctuations themselves were not. 

\begin{figure}[thb]
\centerline{
\includegraphics[width=7.5cm]{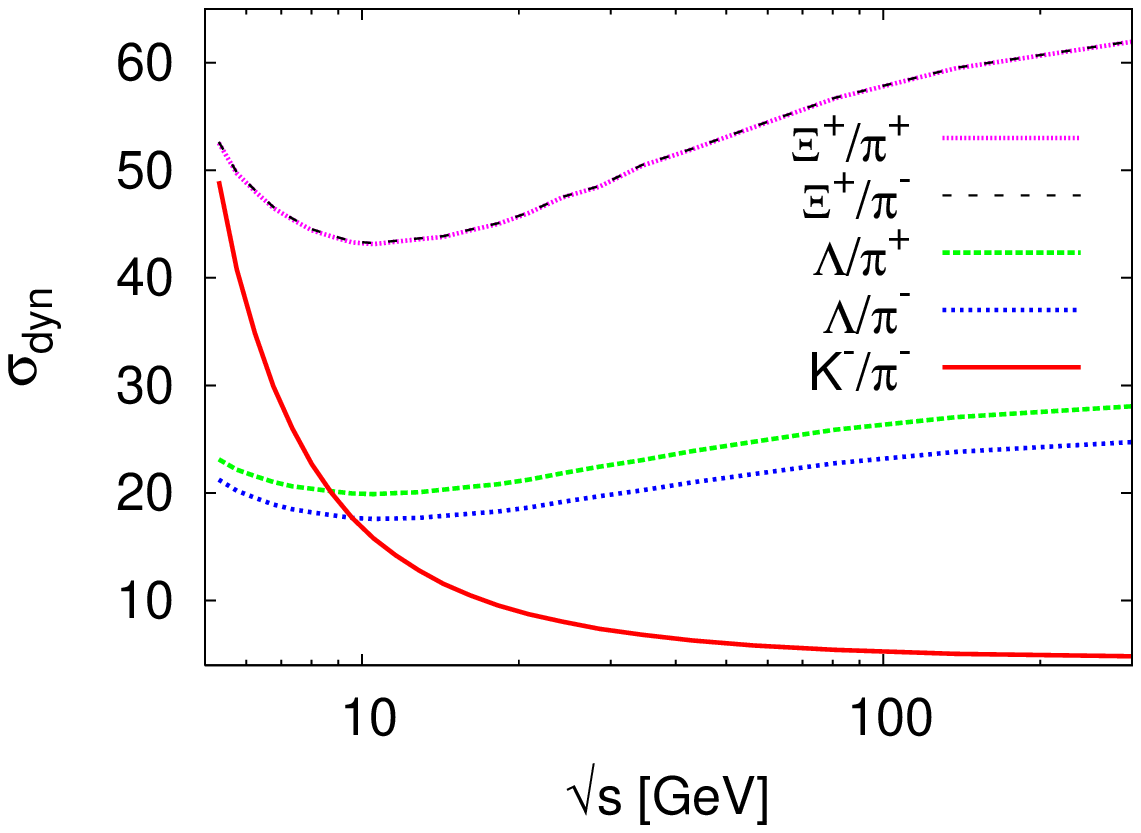}
\includegraphics[width=7.5cm]{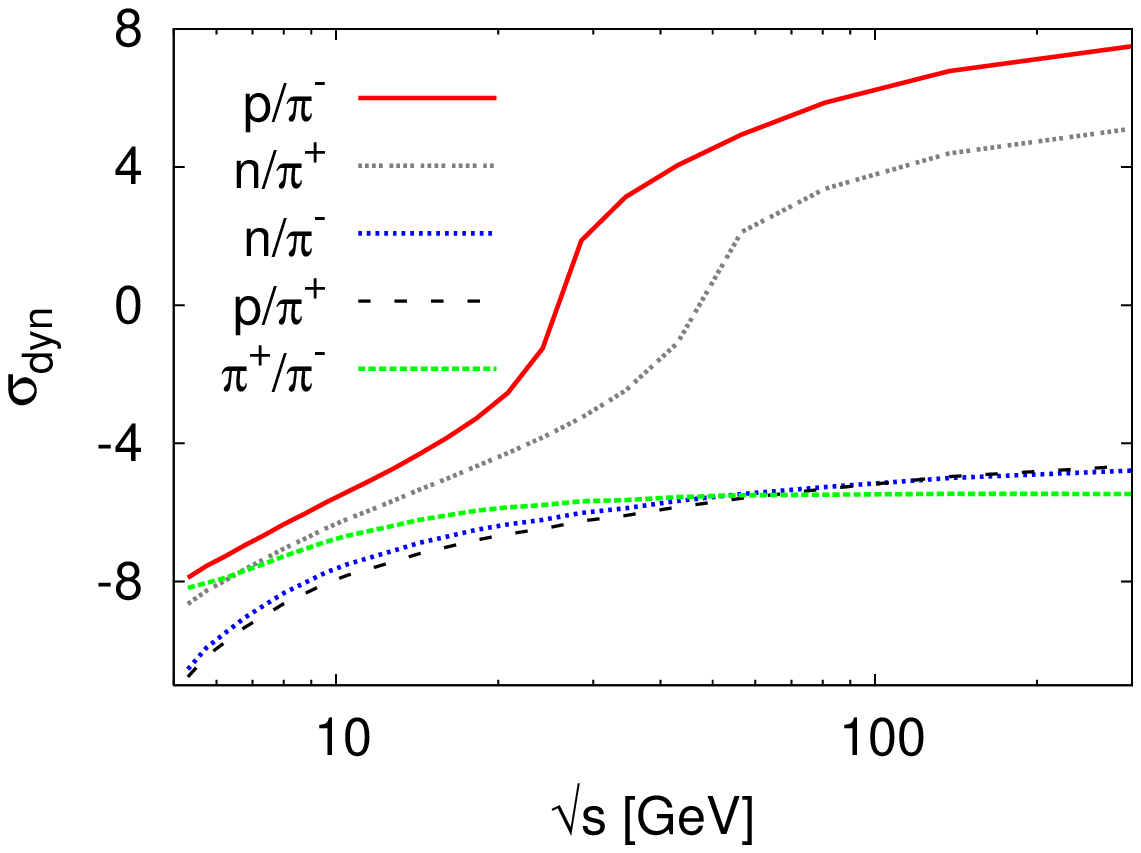}
}
\caption{Predictions for dynamical fluctuations of 
strangeness (left panel) and non-strangeness (right panel) particle
 ratios. Values of dynamical fluctuations at different energies strongly
 depend on particle ratios. } \label{Fig:kmpim}  
\end{figure}

Hadron resonance gas model is obviously
able to reproduce all measurements. For $K^+/\pi^+$ ratio, experimental
data covers a wide range of square root of center of mass 
energy $\sqrt{s}$~\cite{NA492,STAR}. While for
$(p+\bar{p})/(\pi^++\pi^-)$ ratio, 
experimental data available so far  has been measured at SPS energies
only.  That the results from hadron resonance gas model 
agree well with the existing experimental data, allows us to make 
predictions for other particle ratios as will be given in
Fig.~\ref{Fig:kmpim}. On other hand, we can now use the model to
systematically investigate dependence of event-by-event fluctuations on
$\sqrt{s}$.  Few comments are in order at this moment. 
\begin{itemize}
\item Dependence of event-by-event fluctuations of particle ratios on
      $\sqrt{s}$ is non-monotonic
\item Fluctuations  can be suppressed and/or enhanced at different
      $\sqrt{s}$
\item Strangeness fluctuations are positive and enhanced with
      $\sqrt{s}$. Anti-strangeness fluctuations are smaller than
      strangeness ones. There are remarkable minima at top SPS energies
\item Non-strangeness fluctuations are negative, i.e., statistical
      fluctuations are greater than dynamical ones, especially at low energies.
\item In general, event-by-event fluctuations at high energy increase
      with increasing 
      energy. There is only one exception from this empirical
      role; $K^-/\pi^-$ ratio. Fluctuations of
      $K^-/\pi^-$  ratio exponentially decrease with $\sqrt{s}$. 
\end{itemize}

In left panel of Fig.~\ref{Fig:kpi}, we compare experimentally measured
fluctuations of $K^+/\pi^+$ ratios with resonance gas model
results. There is a good agreement at SPS energies. At RHIC, measured
fluctuations are slightly above the theoretical line. The explanation for this
disagreement is
twofolds. First, RHIC measurements are still
preliminary~\cite{STAR}. Second, we refer to our previous study of
particle ratios in heavy ion collisions~\cite{Taw1}. We have concluded
that thermal models with 
$\gamma=1$ slightly overestimate particle ratios at RHIC. To 
reproduce particle ratios, precisely, one has to allow $\gamma$ to take values
different from unity~\footnote[2]{Since available data is still
preliminary, we can, for this moment, restrict our calculations to 
equilibrium value.}. Inserting large  $\langle n\rangle$,  averaged
particle number density, in 
Eq.~\ref{eq:sigma2} apparently results in small $\sigma$.

In right panel in Fig.~\ref{Fig:kpi}, dynamical fluctuations of
$(p+\bar{p})/(\pi^++\pi^-)$ ratios are depicted as a function of
$\sqrt{s}$. According to Eq.~\ref{eq:sigma2},
negative values of dynamical fluctuations are to be understood as a
reason of much dominant statistical fluctuations. Nevertheless, using
our model, we now have a quantitative estimation of both dynamical and
statistical fluctuations. At low SPS energy, there is
a good agreement 
between our model and the experimental results. 

A much more important quantity is the
ratio of dynamical fluctuations to statistical ones. This quantity will
be given in Tab.~\ref{tab:1} at specific energies. Further advantage of
right panel of Fig.~\ref{Fig:kpi} is to illustrate the ability of our model. To
author's knowledge, there is no another experimental data available.

\vspace*{.5cm}

\vspace*{.5cm}

Our predictions for dynamical fluctuations of
different particle ratios are depicted in
Fig.~\ref{Fig:kmpim}. Fluctuations of strangeness particle
ratios are given in left panel. Right panel shows fluctuations of 
non-strangeness particle ratios. Depending on particle ratios, some
fluctuations  decrease and others increase with $\sqrt{s}$. As
we discussed above, the energy dependence is apparently
non-monotonic. Some particle ratios have negative dynamical
fluctuations. Fluctuations of other particle ratios move from negative
to positive values.   

Negative fluctuations are explicitly obtained from non-strangeness particle
ratios; dynamical fluctuations of $n/\pi^-$, $p/\pi^+$ and $\pi^+/\pi^-$
ratios are overall negative. Statistical fluctuations of such 
particle ratios are much greater than dynamical fluctuations.  The
fluctuations of $p/\pi^-$ and $n/\pi^+$ have a remarkable dependence on 
$\sqrt{s}$. At low energy, i.e., AGS and SPS energies, they are
negative. Within a 
relatively short energy interval, their values switch to positive
sign. At higher energies, they monotonically raise with increasing
$\sqrt{s}$. 

The values of dynamical fluctuations of strangeness particle ratios are
greater than that of non-strangeness ones. Their dependence on
$\sqrt{s}$ is also non-monotonic. While fluctuations from $K^-/\pi^-$
exponentially decrease with increasing $\sqrt{s}$, fluctuations from
$\Lambda/\pi$ and $\Xi^+/\pi$ have minimum values at $\sqrt{s}\sim 10\;$GeV. 

Comparing strangeness with non-strangeness fluctuations, we find that 
replacing pion by its anti-particle has almost no influence on
strangeness dynamical fluctuations. Non-strangeness dynamical
fluctuations are dramatically changed, when pion in
denominator~\footnote[1]{Assuming that particle ratios can
mathematically be seen as fractions.} has been replaced by its
anti-particle.  For instance, for $p/\pi$ ratio 
\ba
\langle \left(\Delta n_{p/\pi^+}\right)^2 \rangle &\rightarrow&  
    \langle \left(\Delta n_u\right)^2 \rangle  + 2 \langle \left(\Delta
    n_d\right)^2\rangle  \label{hq1}\\
\langle \left(\Delta n_{p/\pi^-}\right)^2 \rangle &\rightarrow&  
    3 \langle \left(\Delta
    n_u\right)^2\rangle \label{hq2}
\ea   
As in right panel in Fig.~\ref{Fig:kmpim}, while fluctuations of first
    ratio move from negative to positive values, second ratio remains
    negative at all energies.  
In hadronic phase, i.e. particle ratios, quarks are strongly confined
    into hadronic states. It 
    is believed that quarks in the unconfined phase may be strongly
    correlated~\cite{Taw2}. It would be 
    interesting to verify above expressions. In doing this, we have to
    take into consideration volume fluctuations 
    on lattice. Resonance gas model can not be applied at temperatures
    higher than critical one.


\section{\label{sec:4}Discussions and conclusions}

We find that the values of dynamical fluctuations depend on particle
yields.  According to QCD, hadronic matter under extreme conditions of
high temperature and pressure is conjectured to have dynamical
transitions into partonic matter. Assuming that we put certain particle
ratio under these conditions, we want to see the change in its dynamical
fluctuation before and after transition. This is equivalent to our
numerical study at different energies. We start with hadronic matter of two
particles and end  
with partonic mater, Eq.~\ref{hq1} and Eq.~\ref{hq2}. 


Fluctuations of quark number have been studied in lattice
 QCD~\cite{karsch}. It has been found that temperature-dependence
 (equivalent to $\sqrt{s}$-dependence) of fluctuations is dominated by
 analytic part of partition function. Across critical temperature,
 there is a smooth increase of fluctuations with increasing
 temperature. For our analysis, critical temperature is 
 not exactly specified. Our results on fluctuations of individual
 particle are not shown here. Nevertheless, we can make following
 statements: We find that fluctuations of mesons (individual particles) have
 almost same structure as lattice fluctuations and that baryons have maximum
 value at $\sqrt{s}\approx10\;$GeV. At higher energies, baryon
 fluctuations decrease. Quantitative comparison with lattice is - of
 course - not  possible. As 
 discussed in previous section, thermodynamics in hadronic phase can be
 reproduced when taking into account many stable and resonance hadrons in
 the partition function. A free gas of one hadron is not able to reproduce
 the quantitative thermodynamics below critical temperature.  

In Tab.\ref{tab:1}, values of $\sigma_{dyn}/\sigma_{stat}$ quantity of
different particle ratios are
given at specific energies related to certain beam energies at top SPS and
RHIC. It would - of course - be of great interest to verify these values
experimentally. Our predictions specify the values of dynamical
fluctuations related to statistical ones in all heavy ion collisions
experiments. Experimentalists can now decide which particle ratios shall
be measured. It depends - among others - on how large are dynamical
fluctuations compared to statistical ones. 
\begin{table}[htp]
\begin{center}
  \begin{tabular}{|c||c|c|c|c|c|}\hline
       & $12.3$ & $17.3$ & $62.4$ & $100$ & $200$ \\ \hline\hline
$K^+/\pi^+$ & $1.058$ & $1.053$ & $1.080$ & $1.092$& $1.101$ 
   \\ \hline
$K^-/\pi^-$ & $1.381$ & $1.310$ & $1.161$ & $1.146$ & $1.133$ 
   \\ \hline
$\pi^+/\pi^-$ & $0.502$ & $0.458$ & $0.411$ & $0.405$ & $0.403$
   \\ \hline
$(p+\bar{p})/(\pi^++\pi^-)$ & $0.607$ & $0.675$ & $0.803$ & $0.814$
   & $0.821$ \\ \hline
$p/\pi^-$ & $0.851$ & $0.927$ & $1.107$ & $1.139$ & $1.163$ \\
   \hline 
$n/\pi^+$ & $0.778$ & $0.848$ & $1.028$ & $1.059$ & $1.081$ \\
   \hline 
$n/\pi^-$ & $0.605$ & $0.676$ & $0.847$ & $0.873$ & $0.897$ \\
   \hline 
$\Lambda/\pi^+$ & $1.501$ & $1.531$ & $1.573$ & $1.575$ & $1.576$ \\
   \hline 
$\Xi^+/\pi^+$ & $2.367$ & $2.468$ & $2.615$ & $2.625$ & $2.640$ \\
   \hline 
\end{tabular}
  \caption{\label{tab:1}Ratios of $\sigma_{dyn}/\sigma_{stat}$ from
 different particle ratios at specific energies related to known beam
 energies at SPS and RHIC. With this quantity,
 we estimate how large are dynamical fluctuations compared to
 statistical ones at chemical freeze out in different heavy ion
 collisions experiments. Validity of these values depend on validity
 of chemical equilibrium freeze out scenario. }
\end{center}
  \end{table}

Strangeness fluctuation~\cite{marek} are given
in right panels of Fig.~\ref{Fig:kpi} and Fig.~\ref{Fig:kmpim}. The
dependence of strangeness  
fluctuations on energy is also non-monotonic. For $K^+/\pi^+$ ratios,
fluctuations have a negative dependence on energy at SPS. There is a
minimum around $\sqrt{s}\approx20\;$GeV. At higher energies,
fluctuations smoothly increase.  Fluctuations of
$\Lambda/\pi$ and $\Xi^+/\pi$ ratios have also minimum values at $\sqrt{s}\sim
10\;$GeV. $K^-/\pi^-$ ratio
has a completely different behavior. Its dynamical fluctuations of this
particle ratio exponentially decrease with energy.

From our predictions for non-strangeness fluctuations drawn in right panel in
Fig.~\ref{Fig:kmpim}, we find that event-by-event dynamical fluctuations
depend on particle ratios. There is a rapid increase at
low energy. At high energies, fluctuations smoothly increase. There is a
remarkable sharp increase in $p/\pi^-$ and $n/\pi^+$ ratios within a
short range of energies. The values of energies are different. \\

In final conclusion, we propose to study event-by-event dynamical
fluctuations of different particle ratios in heavy ion collisions
experiments. We have shown that baryon to meson fluctuations are much
larger than meson to meson ones. Fluctuations of hybrid and cascade
baryons are much larger than the statistical ones. According to our
results, we now have a framework to study fluctuations of particle ratios,
systematically. As mentioned above, fluctuations of particle ratios
eliminates volume fluctuations. The latter are still included in lattice
simulations. 

We assumed that particle production is due to chemical
equilibrium processes in final state, i.e., $\gamma=1$. 
That our models can very well reproduce experimental measurements means
that the equilibrium freeze out scenario is apparently proved,
especially at SPS energies. At RHIC, precise measurements are needed. 
Energy scan down to $\sqrt{s}=10\;$GeV turns to be a crucial step
to verify the worthwhile behavior of particle
production~\cite{Taw3,Taw1} and now dynamical fluctuations.   

According to this systematic study, it was not possible to point out
certain energy or region of energy, at which dynamical fluctuations
sharply increase. The speculations on manifestation of critical
endpoint by a rapidly increase of dynamical fluctuations of particle
ratios can not be verified by this model. Furthermore, the region of
unconfined phase can not be specified, precisely.  \\

It would be interesting to extract information about the role of
different decay channel in the energy-dependence of dynamical
fluctuations. It will be a further propose to study the effect of
chemical  non-equilibrium  processes of event-by-event dynamical
fluctuations of particle ratios.

\vspace*{10mm}

\noindent
{\bf Acknowledgment}\\ 
This work has been financially supported by the Japanese Society for the
Promotion of Science.

\end{document}